\documentclass[10pt,twocolumn,letterpaper]{article}

\usepackage[pagenumbers]{cvpr} 

\usepackage{graphicx}
\usepackage{amsmath}
\usepackage{amssymb}
\usepackage{makecell}
\usepackage{subcaption}
\usepackage{caption}
\usepackage{multirow}
\usepackage{multicol}
\usepackage[dvipsnames]{xcolor}
\usepackage{soul}
\usepackage{comment}
\usepackage{bbding}
\usepackage{makecell}
\usepackage{xspace}
\usepackage{soul}

\makeatletter
\DeclareRobustCommand\onedot{\futurelet\@let@token\@onedot}
\def\@onedot{\ifx\@let@token.\else.\null\fi\xspace}

\def\eg{\emph{e.g}\onedot} 
\def\ie{\emph{i.e}\onedot} 
 
\def\etc{\emph{etc}\onedot}

\makeatother
\usepackage[normalem]{ulem}
\newcommand{\AY}[1]{{\color{blue} {\bf (AY: #1)}}}%

\newcommand{\tr}[1]{{\color{red} #1}}

\newcommand{\Size}[3]{\mathbb{R}^{#1\times\!#2\times\!#3}}

\newcommand{\imp}[1]{{\color{Black} #1}}
\newcommand{\drp}[1]{{\color{BrickRed} #1}}

\newcommand{\Bst}[1]{\textbf{#1}}
\newcommand{\Scd}[1]{\underline{#1}}
\newcommand{\InsertSubfig}[2]{ \begin{subfigure}[t]{#1\linewidth}
    \centering
        \includegraphics[width=\textwidth]{#2}
    \end{subfigure}}
\newcommand{\InsertSubfigWithCap}[3]{ \begin{subfigure}[t]{#1\linewidth}
    \centering
        \includegraphics[width=\textwidth]{#2}
        \subcaption{#3}
    \end{subfigure}}
\newcommand{\InsertSubfigWithCapWithLabel}[4]{ \begin{subfigure}[t]{#1\linewidth}
    \centering
        \includegraphics[width=\textwidth]{#2}
        \subcaption{#3}
        \label{#4}
    \end{subfigure}}

\definecolor{cvprblue}{rgb}{0.21,0.49,0.74}
\usepackage[pagebackref,breaklinks,colorlinks,citecolor=cvprblue]{hyperref}

\def\paperID{2684} 
\def\confName{CVPR}
\def\confYear{2025}

\title{High-Resolution Be Aware! Improving the 
Self-Supervised \\Real-World Super-Resolution}

\author{Yuehan Zhang \\
National University of Singapore\\
{\tt\small e0546082@u.nus.edu}
\and Angela Yao\\
National University of Singapore\\
{\tt\small ayao@comp.nus.edu.sg}
}

\begin{document}
\maketitle
\begin{abstract}
Self-supervised learning is crucial for super-resolution because ground-truth images are usually unavailable for real-world settings. 
Existing methods derive self-supervision from low-resolution images by creating pseudo-pairs or by enforcing a low-resolution reconstruction objective. These methods struggle with insufficient modeling of real-world degradations and the lack of knowledge about high-resolution imagery, resulting in unnatural super-resolved results.
This paper strengthens awareness of the high-resolution image to improve the self-supervised real-world super-resolution. We propose a controller to adjust the degradation modeling based on the quality of super-resolution results. We also introduce a novel feature-alignment regularizer that directly constrains the distribution of super-resolved images. Our method finetunes the off-the-shelf SR models for a target real-world domain. Experiments show that it produces natural super-resolved images with state-of-the-art perceptual performance.

\end{abstract}
\section{Introduction}

\label{sec:intro}
Image super-resolution (SR) reconstructs high-resolution (HR) images from low-resolution (LR) ones. Real-world LR data contains various other degradations, \eg blur, noise, and compression, with unknown formulations. Due to the cost and difficulty of collecting paired real-world data, supervised methods for real-world SR rely primarily on synthetic data. However, the inevitable domain gaps between synthetic and real-world data lead to unsatisfactory performances. To learn real-world domains directly, self-supervised learning using only LR images is crucial.

\begin{figure}[h!]
    \centering
    \InsertSubfigWithCapWithLabel{0.393}{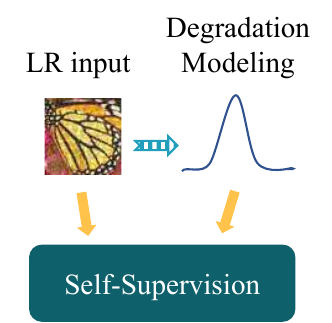}{Previous method}{fig:lway}
    \InsertSubfigWithCapWithLabel{0.558}{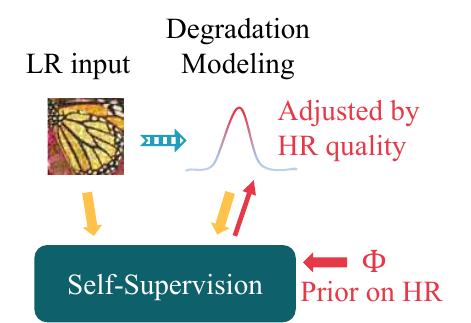}{Ours}{fig:ours}
    \caption{(a) Previous self-supervised SR methods depend on degradation modeling and are limited by only using knowledge of low-resolution (LR) images. (b) Our method strengthens the awareness of high-resolution (HR) images for self-supervision. We adjust the degradation modeling according to the quality of super-resolved HR images and incorporate a prior on HR imagery. 
    }
    \label{fig:teaser}
\end{figure}
Existing self-supervised SR methods use knowledge of LR images in different ways. Many methods~\cite{shocher2018zero,gu2019blind,soh2020meta,deng2023efficient, cheng2020zero} construct pseudo data pairs from the LR inputs by degrading the original LR image to a lower-resolution pseudo-LR input. The original LR image is then used as the super-resolved target, following traditional supervised learning. More recently,~\cite{chen2024low} proposed an LR reconstruction paradigm, which trains an SR model through a pretrained reconstructor aiming at reproducing the LR input from the HR result.

Regardless, the effectiveness of self-supervision with LR images depends on how their real-world degradations are modeled. In the case of pseudo-paired data, the degradations are represented explicitly in the form of degradation kernels. However, the kernels are derived from predefined functions, \eg Gaussian blurring and Gaussian noise, and usually cannot capture the true extent of real-world degradations. This invariably leads to domain gaps of the degradations in pseudo LRs from the original real-world ones.

In the reconstruction paradigm~\cite{chen2024low}, degradations are represented implicitly through degradation embeddings. These embeddings then modulate the reconstructor to degrade the HR images to corresponding LR inputs. Compared to predefined kernels, the high-dimensional embeddings are more informative.
However, if the embedding cannot fully encode the degradation specifics of LR inputs, the SR model is difficult to train effectively. {In such a case, LR reconstruction may even be counter-productive, and the SR model tends to compensate by producing degraded HR outputs.} 

In addition to degradation modeling, a second challenge in self-supervision is that the LR images alone are insufficient to guide the learning toward high-quality super-resolution. Enhancing high-frequency information is essential for SR models to produce natural images~\cite{ledig2017photo,zhou2020guided}, and supervised methods manage to learn high-frequency features from ground-truth HR images~\cite{lin2023catch,zhong2018joint}.
The pseudo-pair methods~\cite{shocher2018zero,gu2019blind,soh2020meta} take original LR inputs, which lack high-frequency features, as the learning target. The LR reconstruction scheme~\cite{chen2024low} calculates loss in the LR space, which has limited effect on high-frequency information lost in the downsampling. Current self-supervised methods struggle to produce natural HR images without introducing knowledge of high-resolution imagery.

Due to insufficient degradation modeling and restricted knowledge from LR images, fulfilling the current self-supervised objectives may not lead to high-quality HR images.
To improve self-supervision for real-world SR, this paper proposes strengthening awareness of high-resolution images.
As shown in \cref{fig:teaser}, we first consider the insufficient degradation modeling and modify the model based on the quality of super-resolved images. We adopt the LR reconstruction scheme~\cite{chen2024low} for degradation modeling, as degradation embeddings are more flexible and informative than predefined kernels. We enrich the degradation embedding with a controller defined by a new quality indicator for the super-resolved images. This adjustment reduces the risk of producing degraded super-resolved images by controlling the degradation extent accordingly.

Second, we propose a novel Feature-Alignment Regularizer (FAR) 
to incorporate prior knowledge from high-resolution imagery. FAR ensures that the characteristics of super-resolved images closely resemble those of natural HR images.
Research has observed that a well-trained self-supervised task shows a significant drop in performance on the test data with shifted distribution~\cite{tsai2023convolutional}.
Inspired by this finding, we design FAR based on a self-supervised task for HR images.
Using an external HR image set, we pretrain an image encoder to align its features with those from an established encoder, \eg CLIP image encoder~\cite{radford2021learning}. 
For the self-supervised learning of the SR model, we enforce this alignment for super-resolved images while keeping the image encoders fixed. This process helps reduce the disparity between the distributions of super-resolved outputs and HR natural images used in the pretraining stage.

{Through the two novel designs, our method adds HR awareness into self-supervised super-resolution. The contributions of this paper are three-fold:
\begin{itemize}
    \item We consider the insufficient degradation modeling and adjust the degradation embedding in the reconstruction paradigm based on the super-resolution quality.
    \item We propose a Feature-Alignment Regularizer for super-resolved images to enhance the natural characteristics.
    \item 
    Our method improves off-the-shelf SR models 
    and achieves state-of-the-art perceptual quality metrics.
\end{itemize}}

\section{Related Works}

\label{sec:related_works}
\noindent\textbf{Real-World Super-Resolution} 
goes beyond standard SR by considering degradations, \eg blur, noise, and compression that accompany low-resolution images in real applications~\cite{cai2019toward}. Collecting paired LR and HR images for real-world settings is challenging and laborious~\cite{cai2019toward,wang2021real}. 
Unsupervised methods~\cite{romero2022unpaired,wei2021unsupervised,zhang2024pairwise} therefore 
use real-world data together with large-scale synthetic datasets. A prevalent data strategy is applying a random set of second-order degradations such as Gaussian noise, Gaussian blur, and JPEG compression, \etc to ground-truth images to synthesize degraded LR images. Such a strategy helps generalize the SR model to unknown real-world degradations and is widely adopted~\cite{wang2021real,zhang2021designing,chen2022real,wang2024exploiting}. However, recent papers observe a generalization-performance trade-off and emphasize optimizing for specific real-world domains~\cite{zhang2024real,zhang2023crafting}.

\noindent\textbf{Self-Supervised Super-Resolution} leverages real-world LR data without heavy training.
ZSSR~\cite{shocher2018zero} constructs LR-HR pairs from the tested LR input by further downsampling, even though this introduces a domain gap from real degradations. KernelGAN~\cite{gu2019blind} estimates blur kernels of LR inputs with an iteratively trained adversarial network. 
SRATT~\cite{deng2023efficient} classifies test data into multiple types of degradations. 
Those methods are restricted by approximating the real degradations through predefined kernels, \eg Gaussian blur, and Gaussian noise. 
Recently, DegAE~\cite{liu2023degae} proposed a novel pretraining paradigm for image enhancement, learning degradation embeddings through an autoencoder. 
LWay~\cite{chen2024low} adapted this strategy for self-supervised real-world SR, using the reconstruction of LR inputs from HR results as a learning objective. The degradation embedding of LR images modulates the reconstructor.

Current self-supervision relies on insufficient degradation modeling and only uses knowledge from LR images.
Our method incorporates awareness of HR images into the self-supervision for adjusting degradation modeling and constraining the distribution of HR images.

\noindent\textbf{Priors for high-resolution} image was used previously in model-based SR formulations~\cite{timofte2015a+,yang2008image,liu2013bayesian}. 
These methods use regularizers on the super-resolved images to constrain characteristics such as sparsity~\cite{timofte2015a+}. Meanwhile, they also enforce a cycle-like consistency between observed LR images and degraded LR samples from super-resolved images under known degradations. 
This formulation was usually within a MAP optimization~\cite{timofte2015a+,liu2013bayesian}. Recent learning-based SR methods also use this formulation and alternatively train the degradation estimation and super-resolution branches~\cite{huang2020unfolding, zhang2020deep}. 
These methods take the learned SR branch, using LR-HR pairs for training, as a prior term. 
In contrast, the prior regularization in our method is for self-supervision and does not require paired data for training.

\section{Preliminaries}
\label{sec:revist}

\subsection{Real-World Super-Resolution}
Consider a super-resolution model $\mathrm{M}$ which restores an HR image $\hat{Y}\in\Size{aH}{aW}{N}$ from the LR input $X\in\Size{H}{W}{N}$, where $a$ is the scaling factor and $N$ is the number of color channels. In supervised SR schemes, paired training data is synthesized by downsampling a ground-truth image $Y^s_{gt}$:
\begin{equation}
X^s = f_{\downarrow}(Y^s_{gt}),
\end{equation}
\noindent to the LR counterpart $X^s$. For standard SR, $f_{\downarrow}$ is a simple and fixed operation such as bicubic downsampling. For real-world SR, $f_{\downarrow}$ is more complex and incorporates degradations other than downsampling. The real-world degradations are complex without known formulations, though they are consistent when arising from the same source, \eg a specific camera. 

Collecting paired real-world datasets is challenging. Existing datasets come from only a limited number of camera sources and the ground-truth images often suffer from blur. 
Consequently, in most real-world SR setups, the LR images $X^r$ are available while the corresponding HR ground truth $Y^r_{gt}$ are assumed absent. 
As such, direct supervised learning for a real-world domain is not feasible.

Many real-world SR methods pursue blind generalization in a supervised manner. They synthesize paired data with complex degradations, \eg using the second-order degradation pipeline~\cite{wang2021real}. Such ``jack-of-all-trades'' models handle diverse input conditions, while specific adaptation for the encountered real-world domain always benefits~\cite{zhang2023crafting,zhang2024real}.
We finetune the generalized SR models for a real-world domain through self-supervised learning. 

\subsection{Self-Supervised Super-Resolution}
\label{sec:SS-SR}

\begin{figure}
    \centering
    \InsertSubfigWithCapWithLabel{0.48}{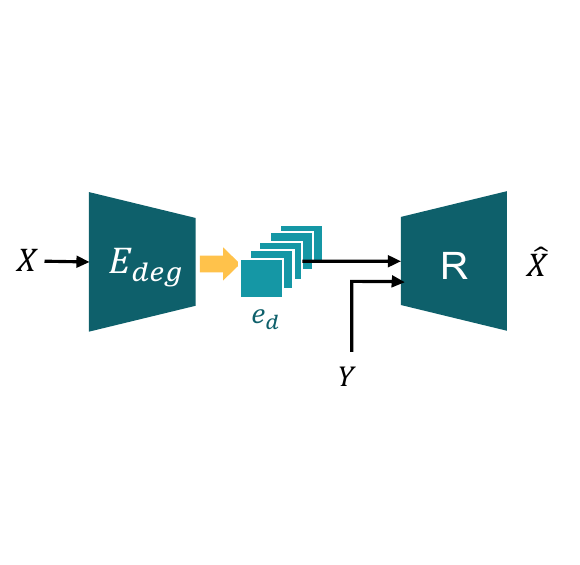}{LR reconstruction network~\cite{chen2024low}}{fig:LRN}
    \InsertSubfigWithCapWithLabel{0.48}{figures/lWay_limi_v2.pdf}{A contradiction phenomenon}{fig:lway_limi}
    \caption{(a) Architecture of LR reconstruction network in~\cite{chen2024low}. (b) The numbers (PSNR$\uparrow$/ LPIPS$\downarrow$) respectively. When the embedding $e_d$ fails to reproduce the blur effect in $X$, passing the ground-truth $Y_{gt}$ through $R$ has lower reconstruction performance than its blurred version. }
\end{figure}

Self-supervised methods~\cite{cheng2020zero,shocher2018zero}  use only real-world LR images ($X^r$) to optimize the SR model, either for a specific test image or for a specific real-world domain with LR images with consistent degradations. 
Previous methods~\cite{gu2019blind,soh2020meta,deng2023efficient, cheng2020zero} using pseudo-paired data approximate the degradation in $X^r$ by predefined kernels and then construct training pairs from $X^r$. 


The LR reconstruction paradigm~\cite{chen2024low} encodes degradations with an embedding learned during pre-training.  
As shown in \cref{fig:LRN}, their LR reconstruction network consists of a degradation encoder $E_{deg}$ and a reconstructor $R$. The reconstructor outputs an LR image $\hat{X}$ based on the degradation embedding $e_d$ and HR image $Y$. 
The LR reconstruction network is first pretrained on an external paired dataset $\{X^s, Y^s_{gt}\}$ with the reconstruction loss $\mathcal{L}_{rec}$:
\begin{equation}
\begin{aligned}
    \theta_{pt}^* &= \arg\min_{\theta_{pt}} \mathcal{L}_{rec}(\hat{X}^s, X^s), \text{ where }\\ &\hat{X}^s = R_{\theta_r}(E_{deg\:\theta_e}(X^s), Y_{gt}^s)
    \label{eq:rec_pt}
\end{aligned}
\end{equation}
where $\theta_{pt}$ consists of parameters $\theta_r$ and $\theta_e$. 

After pretraining, there is a finetuning stage that adapts an off-the-shelf SR model $\mathrm{M}$ to a target real-world domain. This adaptation is done by reconstructing LR samples $X^r$:
\begin{equation}
\begin{aligned}
    \theta^*_{ft} &= \arg\min_{\theta_{ft}} \mathcal{L}_{rec}(\tilde{X}^r, X^r), \text{ where } \\&\tilde{X}^r = R(E_{deg}(X^r), \mathrm{M}_\theta(X^r))
\end{aligned}
\label{eq:rec}
\end{equation}
where the LR reconstruction network is frozen, and $\theta_{ft}$ is from $\theta$, parameters of the SR model $\mathrm{M}$.

The LR reconstruction paradigm (\cref{eq:rec}) assumes the embedding $E_{deg}(X^r)$ can fully encode degradations in $X^r$. As such, reconstructing the LR input $X^r$ in finetuning will lead to high-fidelity $\mathrm{M}(X^r)$ on par with the ground-truth $Y_{gt}^s$ used in pretraining.
However, the embedding can be inadequate for reproducing real-world
degradations in $X^r$ and results in degraded super-resolved images. Besides, without knowledge of high-resolution imagery, the LR reconstruction objective struggles to produce natural HR outputs. 

\section{Methodology}

\begin{figure*}[h!]
\centering
\begin{minipage}{0.90\linewidth}
    \begin{subfigure}[t]{0.48\textwidth}
        \includegraphics[width=\textwidth]{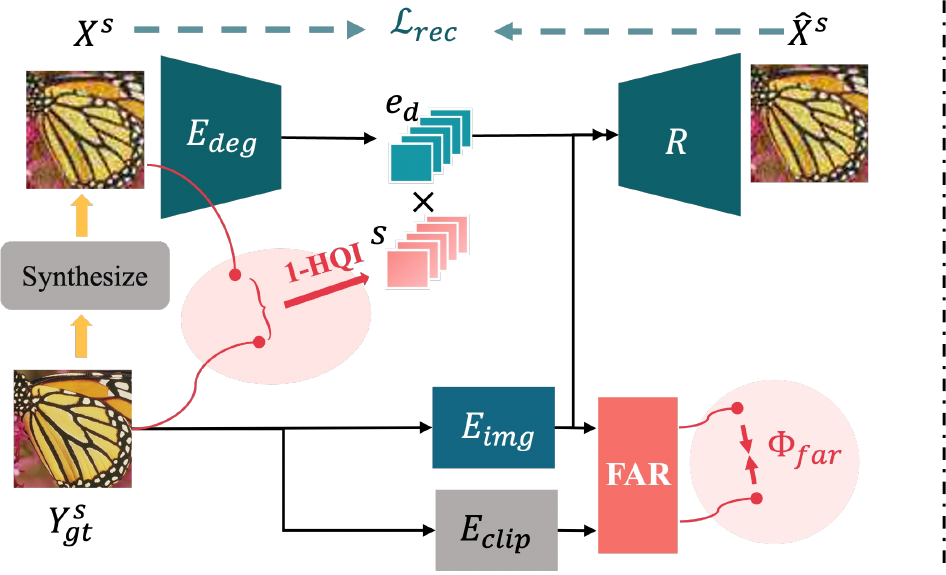}
        \subcaption{Pretraining of LR reconstruction network on synthetic data. }
    \end{subfigure}
    \hfill
    \begin{subfigure}[t]{0.48\textwidth}
        \includegraphics[width=\textwidth]{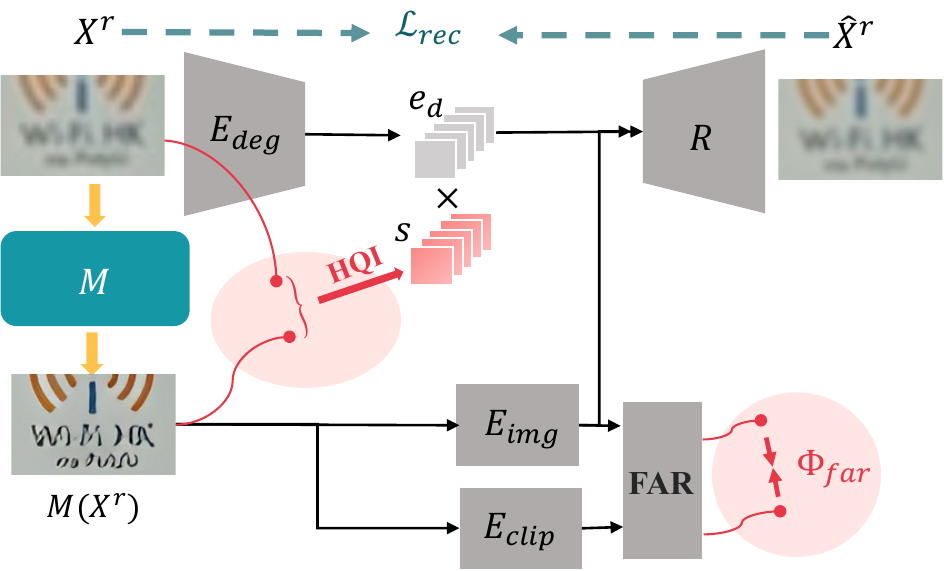}
        \subcaption{Finetuning of SR model $\mathrm{M}$ on real-world data.}
    \end{subfigure}
    \end{minipage}
    \caption{Overview of our method. In each stage, only colored modules are optimized, and LR images are interpolated for better visuals. (a) The pretraining of LR reconstruction network. Parameters in FAR and the reconstruction network ($E_{deg}$, $E_{img}$, and $R$) are optimized with $\mathcal{L}_{rec}$ and $\Phi_{far}$. Controller $\mathbf{s}$ adjusts the degradation embedding $e_d$. LR input $X^s$ is synthesized from HR image $Y^s_{gt}$. 
    (b) Finetuning of super-resolution model $\mathrm{M}$. Given a real-world LR image $X^r$, we input the HR results $\mathrm{M}(X^r)$ to $E_{img}$ and $E_{clip}$. Only parameters in $\mathrm{M}$ are finetuned by $L_{rec}$ and $\Phi_{far}$. The relationships between $\mathbf{s}$ and HQI are different in finetuning and pretraining. 
    }
    \label{fig:overview}

\end{figure*}
\label{sec:methodology}

\cref{fig:overview} is an overview of our method. We follow the reconstruction paradigm~\cite{chen2024low} in utilizing the real-world LR images, as degradation embeddings are more informative and flexible for modification than predefined kernels. 

The key novelty is incorporating HR images in the self-supervision for real-world SR. We adjust the degradation embedding by super-resolution quality and propose a novel feature-alignment regularizer (FAR) for constraining the distribution of super-resolved images.
The pretraining stage on the left optimizes the LR reconstruction network with a novel controller $\mathbf{s}$ by reconstruction loss and the FAR.
The finetuning stage on the right adapts an off-the-self SR model for the test domain with the LR reconstruction network frozen. \cref{subsec:overview} describes our motivation and design in incorporating HR images. \cref{subsec:sys} summarizes the optimizations in the pretraining and finetuning stages.
\cref{subsec:HQI} and \ref{subsec:FAR} explains details of controller $\mathbf{s}$ and FAR respectively. 

\subsection{High-Resolution Awareness}
\label{subsec:overview}

Current self-supervised SR methods rely heavily on degradation modeling and only use the limited information from the LR images. However, the degradation modeling is invariably insufficient to capture real-world degradations. Additionally, it is extremely challenging to learn natural HR outputs solely from the information contained in degraded LR images. To address these limitations, we enhance the awareness of high-resolution images in self-supervision loss by focusing on degradation adjustment and regularization of high-resolution images. 

\noindent \textbf{HR-Aware Degradation Embeddings.} 
In the reconstruction paradigm in Sec.~\ref{sec:SS-SR}, the degradation embedding $e_d$ may not fully capture the degradations in $X^r$. Yet $E_{deg}$ and $R$ are frozen during finetuning, so the only way to compensate for a non-ideal $e_d$ is for the model $\mathrm{M}$ to provide a non-ideal and likely degraded HR image to achieve the reconstruction objective~\cref{eq:rec}. Such a phenomenon is highlighted in the contradiction shown in~\cref{fig:lway_limi}, where passing a ground truth HR image through $R$ leads to lower reconstruction performance than a blurry HR image.

To avoid compromising $M(X^r)$ for LR reconstruction, we design a controller $\mathbf{s}$ to adjust the \emph{extent} of the degradation, which is a primary attribute. 
The controller is generated based on a new High-Resolution Quality Indicator that monitors the quality of HR image $M(X^r)$. For low-quality $M(X^r)$, we expect a greater contribution from $e_d$ to reproduce the degradation and raise the degradation extent accordingly.\\

\noindent \textbf{HR-Aware Image Prior.}
We additionally incorporate extra knowledge about high-resolution natural images. 
Recently,~\cite{tsai2023convolutional} observed that a well-trained self-supervised task performs well on the test data that share the same distribution as the training data but shows a significant drop in the performance when the distribution is shifted.
Inspired by this finding, we assume that a self-supervised task trained with natural HR images can indicate the distribution shift when applied to lower-quality images. Based on this assumption, we have designed our Feature-Alignment Regularizer (FAR) based on a self-supervised task focusing on HR images. 

Specifically, FAR calculates the distance between features extracted by a trainable image encoder $E_{img}$ and an established encoder $E_{clip}$, \ie CLIP image encoder (see \cref{fig:overview}). During pretraining, this distance is minimized for natural HR images ($Y^s_{gt}$). During the finetuning of model $\mathrm{M}$, we minimize such distance for super-resolved image $\mathrm{M}(X^r)$ with fixed image encoders. By pursuing the same alignment performance as pretraining, FAR encourages the distribution of super-resolved results $\mathrm{M}(X^r)$ to be as similar as possible to natural images $Y^s_{gt}$.

\subsection{System Overview}
\label{subsec:sys}
Our approach uses an LR reconstruction network consisting of a degradation encoder $E_{deg}$, image encoder $E_{img}$, and reconstructor $R$.
The reconstruction is based on the embedding $e_d$ and image feature $e_{im}$:
\begin{equation}
\begin{aligned}
    \hat{X} &= R(\mathbf{s}\cdot e_d, e_{im}),\quad\text{where } \\e_d &= E_{deg}(X),\:e_{im} = E_{img}(Y),
\end{aligned}
\label{eq:overview}
\end{equation}
$e_d\in\mathbb{R}^{C_d}$, $e_{im}\in\Size{C_i}{H_i}{W_i}$, and $\mathbf{s}$ is the controller.

As shown in \cref{fig:overview}, the pretraining stage applies reconstruction loss $\mathcal{L}_{rec}$ and regularizer $\Phi_{far}$ to the ground-truth $Y_{gt}^s$ and the synthesized LR image $X^s$:
\begin{equation}
\begin{aligned}
\mathcal{L}_{pt} &= \mathcal{L}_{rec}(R(\mathbf{s}\cdot e_d, e_{im}), X^s) + \lambda_{pt}\Phi_{far}(Y_{gt}^s), \\&\text{where } e_d = E_{deg}(X^s),\:e_{im}=E_{img}(Y_{gt}^s),
\label{eq:overview_pt}
\end{aligned}
\end{equation}
and all parameters, except those from $E_{clip}$, are optimized by minimizing $\mathcal{L}_{pt}$.

In the finetuning stage, losses are calculated for real-world input $X^r$ and the corresponding high-resolution result from an SR model $\mathrm{M}$:
\begin{equation}
\begin{aligned}
\mathcal{L}_{ft} &= \mathcal{L}_{rec}(R(\mathbf{s}\cdot e_d, e_{im}), X^r) + \lambda_{ft}\Phi_{far}(\mathrm{M}(X^r)), \\&\text{where }e_d = E_{deg}(X^r),\:e_{im}=E_{img}(\mathrm{M}(X^r)).
\label{eq:overview_ft}
\end{aligned}
\end{equation}
Only parameters from the SR model $\mathrm{M}$ are optimized by minimizing $\mathcal{L}_{ft}$.

Definition of $\mathcal{L}_{rec}$ follows \cite{chen2024low} and is shown in Supplementary Sec. 8.2.
~\cref{subsec:HQI} and \cref{subsec:FAR} explain the details of controller $\mathbf{s}$ and FAR.

\subsection{Controller s}
\label{subsec:HQI}
As introduced in \cref{subsec:overview}, the specific aim of controller $\mathbf{s}$ is to adjust the degradation extent based on the quality of super-resolved images. 
However, the quality of $\mathrm{M}(X^r)$ cannot be measured directly without corresponding ground truth. 
Given that insufficient degradation embedding is compensated by using degraded HR images to reproduce the LR degradation (see \cref{fig:lway_limi}), 
we evaluate the SR quality by measuring the distance between $X^r$ and $\mathrm{M}(X^r)$. 
A smaller distance between these two images suggests more severe degradations in $\mathrm{M}(X^r)$. 
In practice, we define the High-resolution Quality Indicator (HQI) is defined as:
\begin{equation}
    \text{HQI}(X^r,\mathrm{M}(X^r)) = 1 - \text{LPIPS}(f_\uparrow(X^r), \mathrm{M}(X^r)), 
\end{equation}
where $f_\uparrow$ is bicubic upsampling to align spatial resolution of $X^r$ to $\mathrm{M}(X^r)$, and LPIPS~\cite{zhang2018unreasonable} calculates the cosince \textit{distance} $d\in [0,1]$ between image features. 

In the finetuning stage, when $X^r$ and $\mathrm{M}(X^r)$ are close, \ie \textit{high} HQI, we raise the degradation extent to lower the risk of degrading the HR image for the reconstruction of the LR input.
In the pretraining stage, as the HR image $Y^s_{gt}$ is ground-truth, the similarity between $X^s$ and $Y^s_{gt}$ reflects the "ground-truth'' degradation extent, \ie a close distance should correlate to low degradation extent. 
The controller $\mathbf{s}$ in \cref{eq:overview} is thus defined differently for the pretraining and finetuning stages:
\begin{equation}
 \mathbf{s} =\begin{cases}
    \mathbf{n} + (1-\text{HQI}(X^s, Y^s_{gt})),  & \text{pretraining}\\
    \mathbf{n} + \text{HQI}(X^r, \mathrm{M}(X^r)),   & \text{finetuning}
\end{cases},
\label{eq:s}
\end{equation}
where the vector $\mathbf{n}$ is a $C_d$-dimensional random vector and $\mathbf{n} \sim \mathcal{N}(\mathbf{0}, \mathbf{I})$, which helps the robustness.

\subsection{Feature-Alignment Regularizer (FAR)}
\label{subsec:FAR}
\begin{figure}
    \centering
    \includegraphics[width=0.85\linewidth]{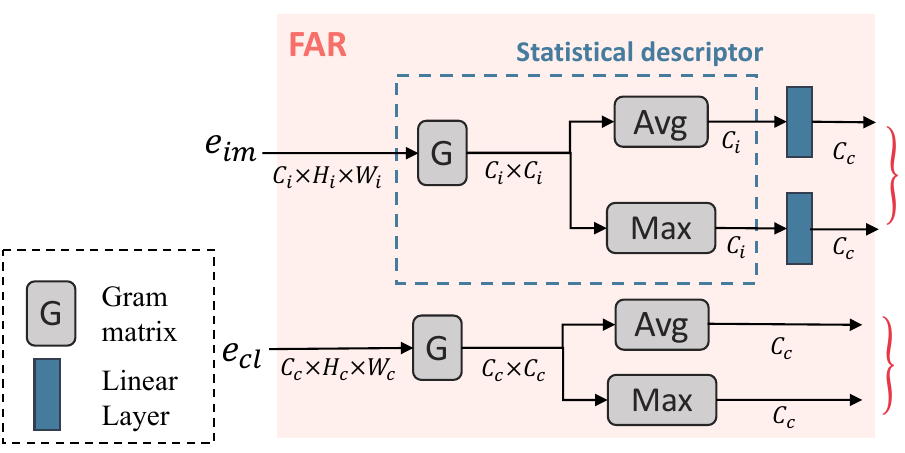}
    \caption{Details of the Feature-Alignment Regularizer. }
    \label{fig:far}
\end{figure}

FAR constrains the distribution of super-resolved image $\mathrm{M}(X^r)$ so that they exhibit natural characteristics that cannot be learned from LR images. FAR calculates a self-supervision loss for a HR image $Y$, aiming to align the features of $Y$ encoded by $E_{img}$ and those processed by the established image encoder $E_{clip}$ from CLIP~\cite{radford2021learning} model:
\begin{equation}
\begin{aligned}
    \Phi_{far}(Y) = \mathcal{L}_{far}(e_{im}, e_{cl}), \quad\text{and}\\ e_{im}=E_{img}(Y),\: e_{cl} = E_{clip}(Y), 
\end{aligned}
\label{eq:im_feat}
\end{equation}
where $e_{im}\in\Size{C_i}{H_i}{W_i}$ and $e_{cl}\in\Size{C_c}{H_c}{W_c}$.

$\mathcal{L}_{far}$ emphasizes statistical differences between feature maps, allowing loss values to better reflect distribution shifts.
As in \cref{fig:far}, we design a statistical descriptor based on the Gram Matrix~\cite{gatys2016image} for an image feature $e\in\Size{C}{H}{W}$: 
\begin{equation}
\begin{aligned}
    &S_{avg}(e) = \text{AvgPool}(G(e)), \\&S_{max}(e) = \text{MaxPool}(G(e)),
\end{aligned}
\label{eq:stat_des}
\end{equation}
where $G$ calculates the Gram Matrix, \ie channel correlation of feature maps, and  $G(e)\in\mathbb{R}^{C\times\!C}$. The Gram matrix of a deep feature focuses on the statistical features rather than image content~\cite{gatys2016image}.
We use the average and maximum values of each channel's correlation to others as the statistical descriptor, and both $S_{avg}$ and $S_{max}$ are of size $\mathbb{R}^C$. 

We then define $\mathcal{L}_{far}$ as difference between statistical descriptors of $e_{im}$ and $e_{cl}$ in \cref{eq:im_feat}. As $e_{im}$ and $e_{cl}$ usually do not have the same channel number, \ie $C_i\neq C_c$, we use Linear layers to align the dimensions between the two descriptors: 
\begin{equation}
    \begin{aligned}
        \mathcal{L}_{far}&(e_{im}, e_{cl}) = \|\mathbb{T}_a\cdot S_{avg}(e_{im}) - S_{avg}(e_{cl})\|_\mathcal{F}\\
    &+ \|\mathbb{T}_m\cdot S_{max}(e_{im}) - S_{max}(e_{cl})\|_\mathcal{F},
    \end{aligned}
    \label{eq:far}
\end{equation}
where $\mathbb{T}_a$ and $\mathbb{T}_m$ are Linear layers learnable only in the pretraining stage, and $\|\cdot\|_\mathcal{F}$ is the Frobenius norm.

{The pretraining stage minimizes $\Phi_{far}(Y^s_{gt})$ to optimize the alignment relationship on ground-truth HR images of the synthetic dataset, which are of natural image distribution. During finetuning of $\mathrm{M}$, we minimize $\Phi_{far}(\mathrm{M}(X^r))$ to constraint the disparity of distribution between $\mathrm{M}(X^r)$ and the natural ground-truth images. }

\begin{table*}[h!]
    \centering
    \resizebox{\linewidth}{!}{
    \begin{tabular}{c|c|cc|cccc|cccc}
    \hline
    \hline
       {Datasets} & {Metrics}&\makecell[c]{ZSSR\\~\cite{shocher2018zero}} &\makecell[c]{KernelGAN\cite{gu2019blind}\\+ZSSR}& \makecell[c]{RealESRGAN+\\~\cite{wang2021real}}&\makecell[c]{+SRTTA~\cite{deng2023efficient}} & +LWay~\cite{chen2024low} &+Ours&SwinIRGAN&+SRTTA &+LWay &+Ours\\
        \hline
        \multirow{4}{*}{NTIRE20}&PSNR$\uparrow$&26.11 &16.20 &25.08 &\imp{{25.51}} & \imp{25.41}& \imp{{25.87}} & 26.93&\imp{{27.09}} &\imp{26.96} &\imp{{27.04}} \\
        &SSIM$\uparrow$&0.6841 &0.2777 &0.7061 &\imp{{0.7212}} &\imp{0.7142} & \imp{{0.7204}} &0.7467 & \imp{{0.7513}}&\imp{0.7474} & \imp{{0.7497}} \\
        &LPIPS$\downarrow$&0.6465 & 0.7446&0.2504 &\drp{0.2718} &\imp{\Scd{0.2498}} & \imp{\Bst{0.2369}} &0.2290& \drp{0.2417} &\imp{\Scd{0.2232}} & \imp{\Bst{0.2178}} \\
        &NRQM$\uparrow$&4.612 &5.019 &\Scd{6.121} & \drp{5.885} & \drp{6.022} & \imp{\Bst{6.137}} &\Scd{5.740} & \drp{5.476} & \drp{5.734} & \imp{\Bst{5.804}} \\
        \hline
        \hline
        \multirow{4}{*}{AIM2019}&PSNR$\uparrow$&22.84 & 16.29&22.58&\imp{{23.06}} &\imp{{22.77}} &\imp{22.72} &23.14&\imp{{23.22}} &\imp{{23.15}} &{23.13} \\
        &SSIM$\uparrow$&0.6233 &0.3335 &0.6444&\imp{{0.6686}} &\imp{{0.6525}} &\imp{0.6470} &0.6634 &\imp{{0.6692}}&\imp{{0.6667}} & \imp{0.6652}  \\
        &LPIPS$\downarrow$&0.6446 &0.7547 &0.2959 &\drp{0.3459} &\imp{\Scd{0.2942}} & \imp{\Bst{0.2894}} &0.2897& \drp{0.3145} & \imp{\Scd{0.2871}} & \imp{\Bst{0.2831}}  \\
        &NRQM$\uparrow$&3.765 &4.577 &\Scd{6.243} & \drp{5.357} & \drp{6.077} & \imp{\Bst{6.252}} & \Scd{5.959} & \drp{5.491} & \drp{5.885} & \imp{\Bst{5.961}} \\
        \hline
        \hline
        \multirow{4}{*}{\makecell[c]{RealSR-\\Canon}}&PSNR$\uparrow$&26.45 &23.06 &24.74 & \imp{24.97}&\imp{{25.77}} &\imp{{25.46}} &{26.77} &\imp{{26.78}} &\drp{26.74} &\drp{26.52} \\
        &SSIM$\uparrow$&0.7631 &0.6385 &0.7634 &\imp{{0.7788}} &\imp{{0.785}} &\imp{0.7674} &0.7892 &\imp{{0.7898}} &\imp{{0.7896}} &\imp{0.7895} \\
        &LPIPS$\downarrow$&0.3999 &0.4506 &0.2607 & \imp{\Scd{0.2470}}&\imp{\Bst{0.2460}} &\imp{0.2510} &0.2576 &\imp{\Scd{0.2513}} & \imp{0.2567} & \imp{\Bst{0.2506}}  \\
        &NRQM$\uparrow$&2.861 &4.0526 &\Scd{6.065} & \drp{5.438} & \drp{5.118} &\imp{\Bst{6.098}} & \Scd{4.225} & \drp{4.169} & \drp{4.193} & \imp{\Bst{4.335}}  \\
        \hline
        \multirow{4}{*}{\makecell[c]{RealSR-\\Nikon}}&PSNR$\uparrow$&26.09 &21.34 &24.31 & \drp{24.07}&\imp{{25.21}} &\imp{{25.14}} & {26.25} & \imp{{26.25}} &\drp{26.20} &\drp{25.97} \\
        &SSIM$\uparrow$&0.7378 &0.5871  &0.7334 &\imp{0.7399} & \imp{{0.7532}} & \imp{{0.7449}} &0.7602 &\imp{{0.7615}} & \imp{0.7607} & \imp{{0.7607}} \\
        &LPIPS$\downarrow$&0.4114 & 0.4619 &0.2851 & \imp{0.2796}& \imp{\Scd{0.2781}} & \imp{\Bst{0.2681}} &0.2796 & \imp{\Scd{0.2758}} & \imp{0.2785} & \imp{\Bst{0.2729}}\\
        &NRQM$\uparrow$&3.156&4.385 &\Scd{5.669} & \drp{5.334} & \drp{5.163} & \imp{\Bst{5.739}} &\Scd{4.854} & \drp{4.799} & \drp{4.819} & \imp{\Bst{4.889}} \\
        \hline
        \hline
    \end{tabular}}
    \caption{Quantitative comparisons. \drp{Red} colors highlight the \drp{drop} compared to the off-the-shelf SR model for finetuning. \textbf{Bold} and \underline{underline} emphasize the best and second-best perceptual metrics among methods for finetuning. Our method achieves state-of-the-art finetuning performance for the essential perceptual quality.
    }
    \vspace{-0.2cm}
    \label{tab:quant_cmp}
\end{table*}
\section{Experiments}
\label{sec:experiments}
\subsection{Implementations}
\label{subsec:implementation}

\noindent\textbf{Architectures.} $E_{img}$ is a lightweight convolutional network, consisting of six residual blocks of 64-channel internal features. $E_{img}$ does not reduce the spatial resolution of features.
$E_{deg}$ is also residual and composed of 16 residual blocks. The output feature is a 512-dimensional vector. The design of reconstructor $R$ combines SRResNet~\cite{ledig2017photo} and StyleGAN2~\cite{karras2020analyzing}. We use the internal feature of the pretrained CLIP ``RN50" model as the output from $E_{clip}$. Other details are in Supplementary Sec. 8.1. 

\noindent\textbf{Important Training Configurations.}
The LR reconstruction network is pretrained with a learning rate of $2e\!-\!4$ with a cosine annealing schedule~\cite{loshchilov2016sgdr} 
for 500K iterations, using the Adam optimizer~\cite{kingma2014adam}, with 
$(\beta_1,\beta_2)\!=\!(0.9, 0.999)$. 
Images are cropped into 
$64\!\times\!64$ patches. 
For finetuning, the learning rate is set to $2e\!-\!6 \sim 5e\!-\!6$, depending on the architecture of SR models. The SR models are 
finetuned for 300-600 iterations. 
We use the Exponential Moving Average strategy with a decay rate of 0.999 in both the pretraining and finetuning stages. Other details of the pretraining and finetuning are given in Supplementary Sec. 8.3.

\noindent\textbf{Datasets and Evaluations.} For the pretraining, we use the ground-truth images from the training set of DIV2K~\cite{agustsson2017ntire} and synthesize LR counterparts through the second-order degradation pipeline~\cite{wang2021real}. 
We finetune off-the-shelf SR models with our methods on the testing sets of NTIRE20~\cite{lugmayr2020ntire}, AIM2019~\cite{lugmayr2019aim}, and RealSR~\cite{cai2019toward} respectively. The finetuning is on the whole testing set, which has consistent degradation, for the computation-saving aim.  
These datasets have unknown degradations from synthetic or real-world sources and provide paired HR images for evaluation. 
The super-resolved images are evaluated with
PSNR, SSIM, LPIPS~\cite{zhang2018unreasonable}, and NRQM~\cite{ma2017learning}. The first three metrics are full-reference. PSNR is measured on the Y channel of images in YCrCb format. NRQM is a no-reference metric focusing on statistical characteristics, such as the clearness, of the images. PSNR and SSIM evaluate pixel-wise fidelity, while LPIPS and NRQM focus on perceptual quality.

\subsection{Comparisons with Other Methods}
\noindent\textbf{Quantitative comparisons} are between our methods and ZSSR\cite{shocher2018zero}, KernalGAN\cite{gu2019blind}+ZSSR, SRTTA\cite{deng2023efficient}, and LWay\cite{chen2024low}. LWay does not release source code and their self-collected dataset for pretraining the LR reconstruction network. We reproduce their method based on the official instructions and use the same synthetic data in \cref{subsec:implementation} for pretraining. Supplementary Sec. 7 explains the details of reproduction. 
We apply SRTTA, LWay, and our method to finetune 
off-the-shelf SR models RealESRGAN+~\cite{wang2021real} and SwinIRGAN. SwinIRGAN is built by training SwinIR~\cite{liang2021swinir} with second-order degradation~\cite{wang2021real} and an adversarial loss. 
As shown in \cref{tab:quant_cmp}, ZSSR has poor perceptual scores, and using KernelGAN with it improves image sharpness (NRQM) but leads to poor fidelity (PSNR and SSIM). 
SRTTA improves the pixel-wise fidelity compared to original SR models but always drops in perceptual quality. Similarly, LWay constantly decreases NRQM across datasets. Instead, our method improves SR models for both fidelity and perceptual qualities and achieves the best improvements on the important perceptual evaluations.

\begin{figure*}[h!]
\centering
    \InsertSubfig{0.125}{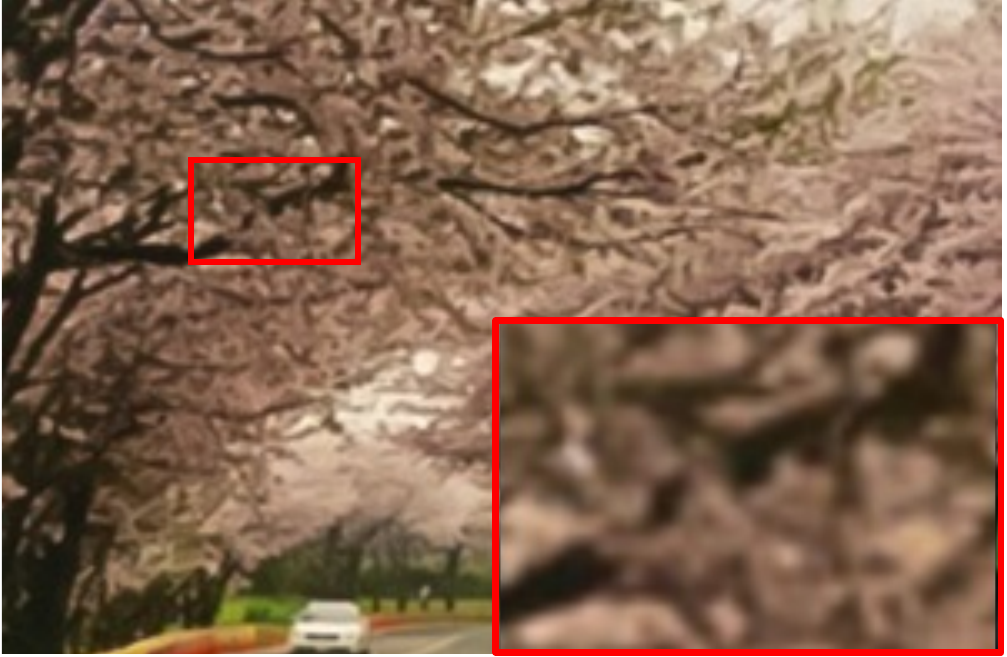}
    \InsertSubfig{0.125}{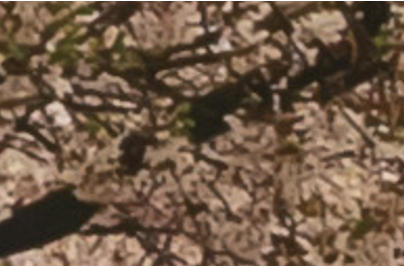}
    \InsertSubfig{0.125}{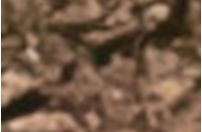}
    \InsertSubfig{0.125}{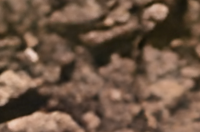}
    \InsertSubfig{0.125}{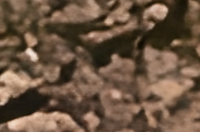}
    \InsertSubfig{0.125}{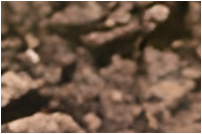}
    \InsertSubfig{0.125}{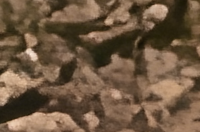}

    \InsertSubfig{0.125}{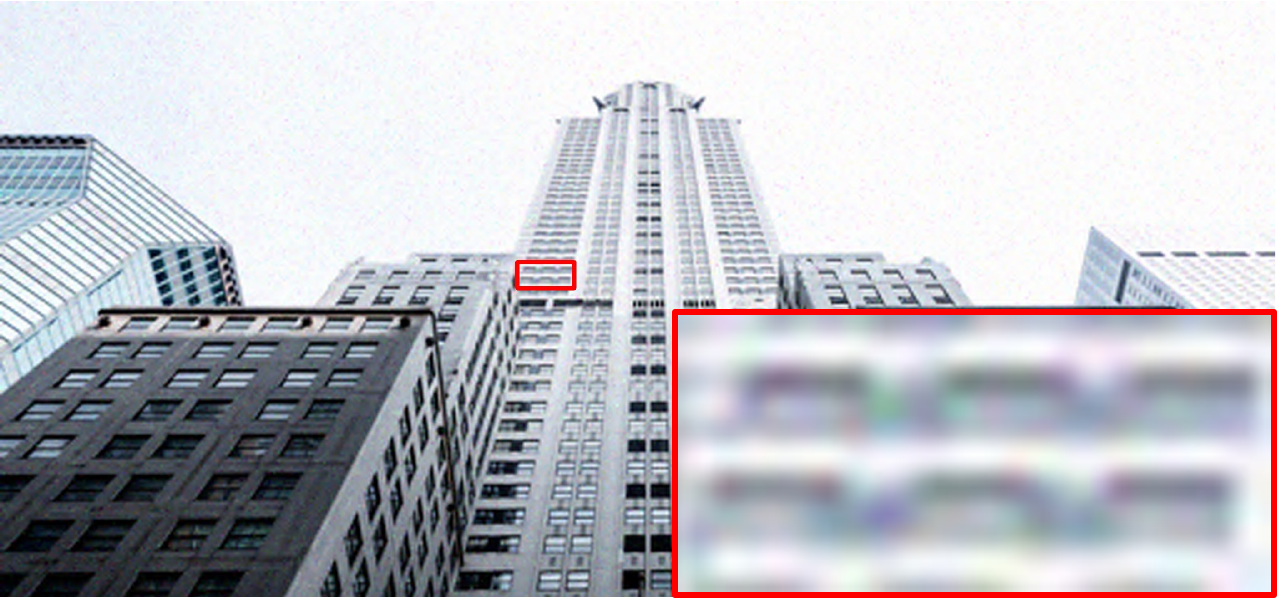}
    \InsertSubfig{0.125}{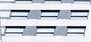}
    \InsertSubfig{0.125}{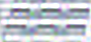}
    \InsertSubfig{0.125}{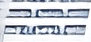}
    \InsertSubfig{0.125}{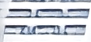}
    \InsertSubfig{0.125}{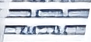}
    \InsertSubfig{0.125}{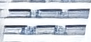}

    \InsertSubfig{0.125}{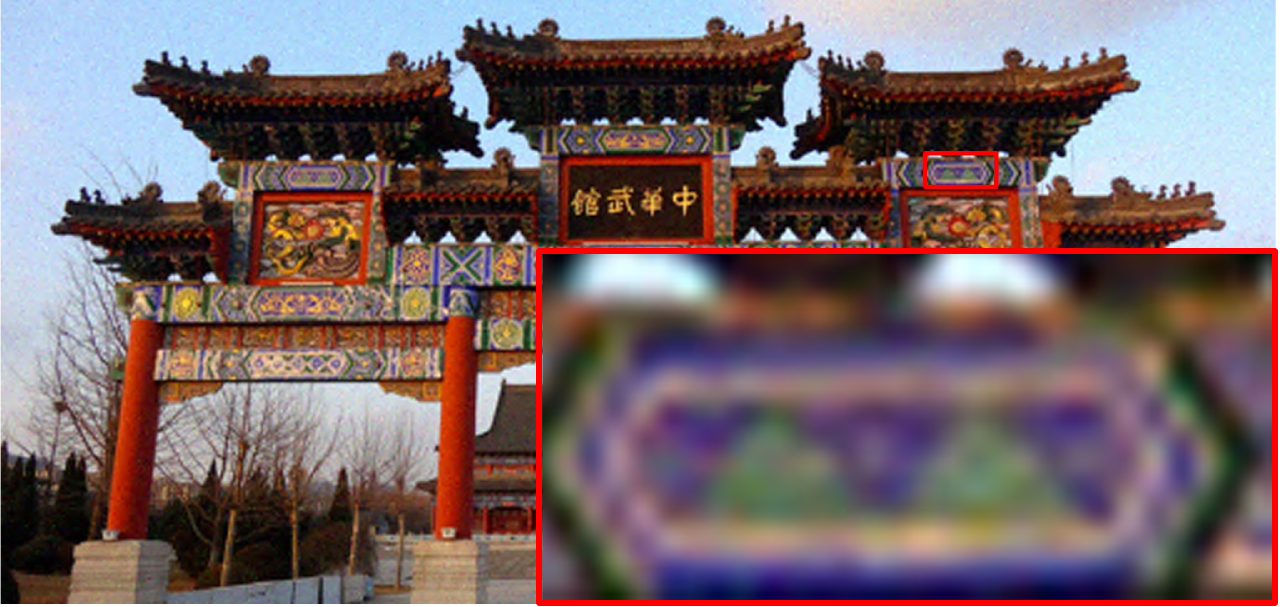}
    \InsertSubfig{0.125}{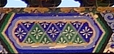}
    \InsertSubfig{0.125}{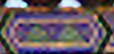}
    \InsertSubfig{0.125}{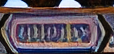}
    \InsertSubfig{0.125}{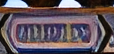}
    \InsertSubfig{0.125}{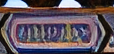}
    \InsertSubfig{0.125}{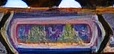}


    \InsertSubfigWithCap{0.125}{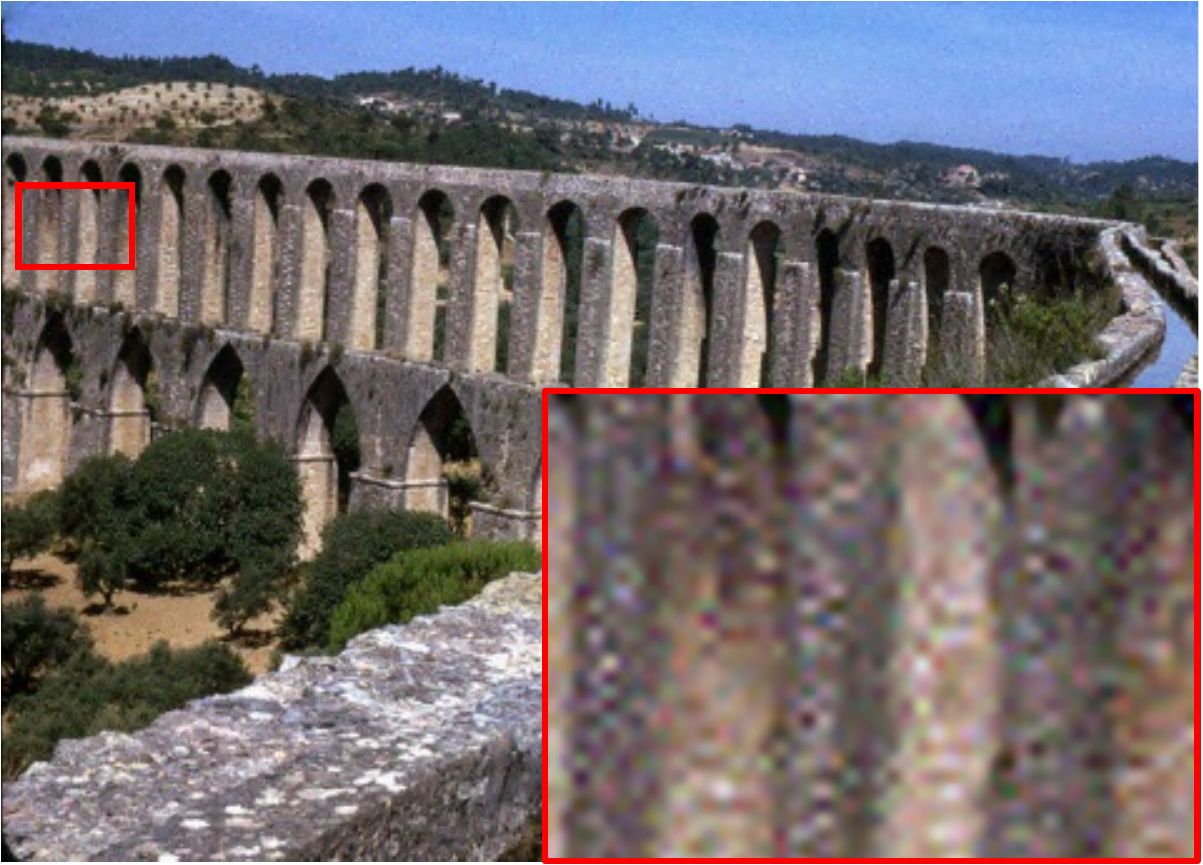}{LR}
    \InsertSubfigWithCap{0.125}{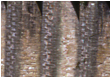}{HR}
    \InsertSubfigWithCap{0.125}{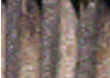}{ZSSR}
    \InsertSubfigWithCap{0.125}{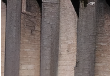}{RE}
    \InsertSubfigWithCap{0.125}{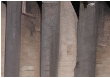}{+SRTTA}
    \InsertSubfigWithCap{0.125}{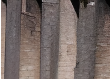}{+LWay}
    \InsertSubfigWithCap{0.125}{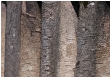}{+Ours}
    \vspace{-0.2cm}
    \caption{Visual comparison with state-of-the-art methods (zoom-in for better views). RE refers to RealESRGAN+ and columns (e-g) are the results of methods finetuning RE. Compared to SRTTA~\cite{deng2023efficient} and LWay~\cite{chen2024low}, our method restores shapes close to HR images, such as windows in the second row and triangles in the third row, and generates realistic patterns.}
    \vspace{-0.2cm}
    \label{fig:cmp_sota}
\end{figure*}

\begin{figure}[h!]
\centering
\begin{minipage}{0.95\linewidth}
    \centering
    \InsertSubfig{0.23}{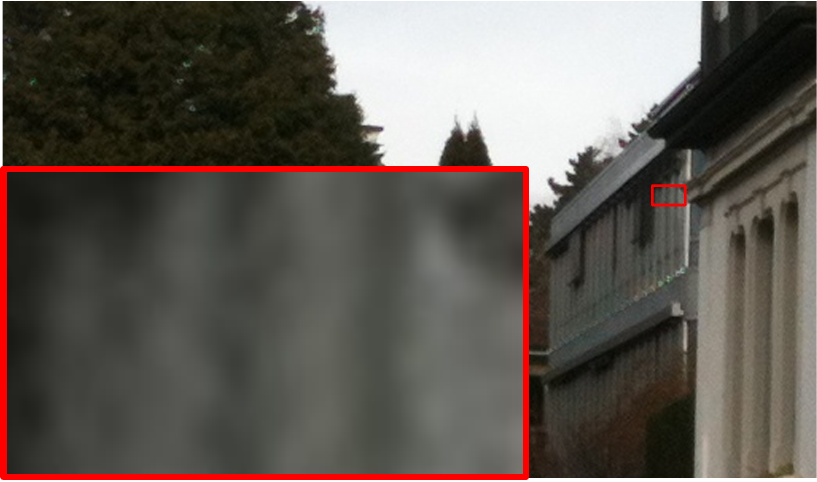}
    \InsertSubfig{0.23}{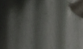}
    \InsertSubfig{0.23}{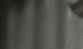}
    \InsertSubfig{0.23}{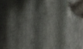}

    \InsertSubfig{0.23}{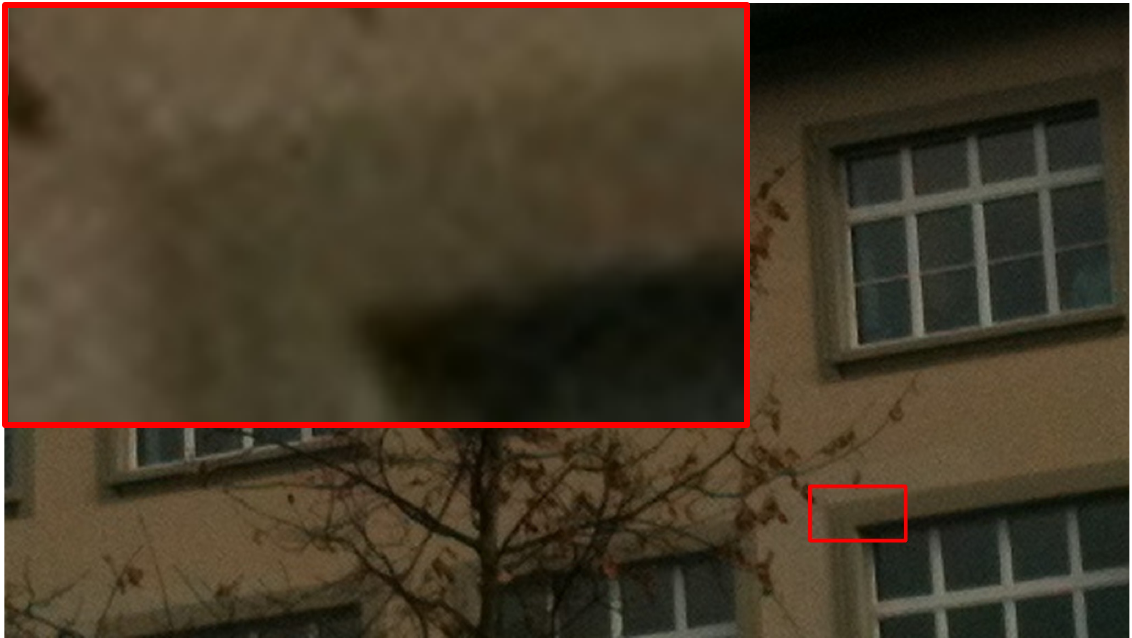}
    \InsertSubfig{0.23}{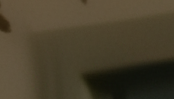}
    \InsertSubfig{0.23}{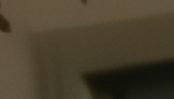}
    \InsertSubfig{0.23}{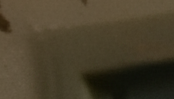}
    
    \InsertSubfigWithCap{0.23}{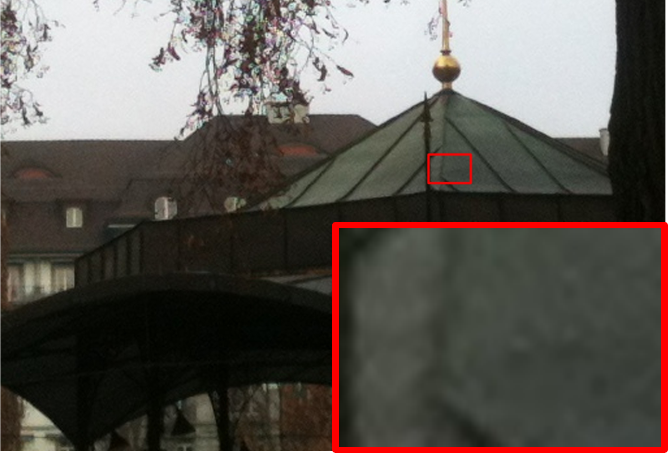}{LR}
    \InsertSubfigWithCap{0.23}{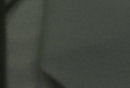}{Swin}
    \InsertSubfigWithCap{0.23}{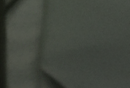}{+LWay}
    \InsertSubfigWithCap{0.23}{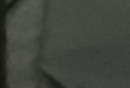}{+Ours}
\end{minipage}
\vspace{-0.2cm}
    \caption{Visual results for DPED~\cite{ignatov2017dslr} dataset. Methods in (c-d) finetune SwinIRGAN (Swin). LWay fails to improve the over-smoothness produced by Swin. Our method reconstructs realistic textures that better resemble the original images.  }
    \label{fig:cmp_dped}
\end{figure}

\noindent\textbf{Qualitative comparisons} is shown in \cref{fig:cmp_sota}. ZSSR yields blurry results. In comparison to RealESRGAN+ and models finetuned using SRTTA and LWay, our method super-resolves sharper and more realistic patterns similar to the HR image. \cref{fig:cmp_dped} shows results for a real-world dataset without ground-truth images. SwinIRGAN (Swin) and the model tuned with LWay produce over-smoothed patterns. Instead, our method generates realistic textures.
\subsection{Ablations}
\noindent\textbf{Effectiveness of Designs.} \cref{tab:abla_func} shows ablation results of the controller $\mathbf{s}$ and the FAR regularizer by finetuning the RealESRGAN+ model. 
The baseline method only uses $\mathcal{L}_{rec}$ without the controller $\mathbf{s}$ in the pretraining and finetuning stages.
The baseline performs well only on pixel fidelity. Adding either the controller $\mathbf{s}$ or FAR improves the perceptual quality (LPIPS and NRQM); of the two, adjusting the degradation extent favors LPIPS
more than NRQM. Combining the two strategies achieves balanced improvements in all metrics over the off-the-shelf SR model.

\begin{table}[h!]
    \centering
    \resizebox{\linewidth}{!}{
    \begin{tabular}{c c | c c c c}
    \hline
    \hline
    \multicolumn{2}{c|}{Method} & \multicolumn{4}{c}{Metrics}\\
    \hline
      controller $s$  & FAR & PSNR$\uparrow$ & SSIM$\uparrow$ & LPIPS$\downarrow$ & NRQM$\uparrow$ \\
    \hline
      \XSolidBrush & \XSolidBrush &26.2 &0.7889 &0.2525 &4.511 \\
      \Checkmark & \XSolidBrush &{25.78} &{0.7858} &{0.2434} &4.979 \\
      \XSolidBrush & \Checkmark &25.4 &0.7668 &0.2522 & {6.113} \\
      \Checkmark & \Checkmark &{25.46} &{0.7674} &{0.251} &{6.099} \\
      
    \hline
    \hline
    \end{tabular}}
    \vspace{-0.1cm}
    \caption{Ablation on the controller $\mathbf{s}$ and the regularizer FAR. Compared to the baseline (first row), using either $\mathbf{s}$ or FAR achieves better perceptual scores. Using two strategies together yields balanced improvements over the pretrained SR model. 
    }
    \vspace{-0.2cm}
    \label{tab:abla_func}
\end{table}

\noindent\textbf{LR reconstruction vs. SR performance.}
The phenomenon illustrated in \cref{fig:lway_limi} shows that an HR image with the same type of degradation as the LR input can yield a higher reconstruction performance than the ground-truth image. 
This motivates our design of controller $\mathbf{s}$ to mitigate the risk of degrading HR images for LR reconstruction performance.
%
Here, we show the effect of $\mathbf{s}$ on the reconstruction objective. We interpolate the bicubic-upsampled ($f_\uparrow$) LR input and ground-truth HR image to create HR images with different extents of LR degradation:
\begin{equation}
    Y_i = i \cdot f_\uparrow(X) + (1-i) \cdot Y_{gt},
    \label{eq:y_i}
\end{equation}
where $i\in[0,1]$ is the ratio of interpolation. 

\begin{figure}
    \centering
    \InsertSubfig{0.475}{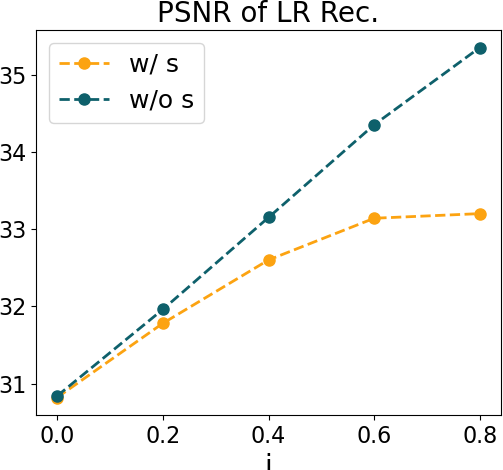}
    \InsertSubfig{0.485}{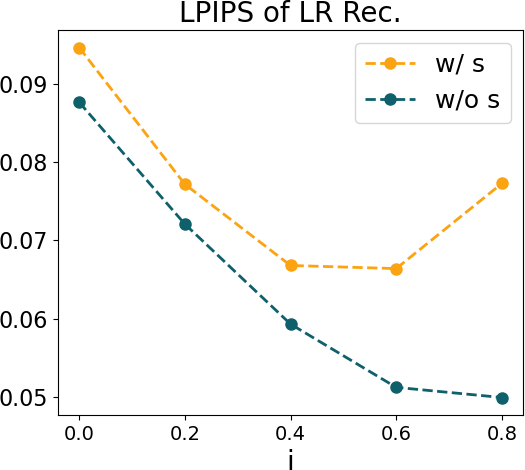}
    \caption{Plot of LR reconstruction performance changing with the interpolation ratio $i$ in \cref{eq:y_i}.}
    \label{fig:abla_rec}
\end{figure}

In \cref{fig:abla_rec}, we present the PSNR and LPIPS scores comparing $X$ and the reconstructed $\hat{X}$. The reconstruction uses \cref{eq:overview} with different $Y_i$. We train two LR reconstruction networks distinguished by using controller $\mathbf{s}$ or not. Neither network utilizes FAR to eliminate external factors. 
As shown in \cref{fig:abla_rec}, without $\mathbf{s}$, lower quality $Y_i$ constantly yields higher reconstruction performance. The controller $\mathbf{s}$ stops this trend, discouraging SR models from degrading HR images for better LR reconstruction. 

\begin{figure}
    \centering
\includegraphics[width=\linewidth]{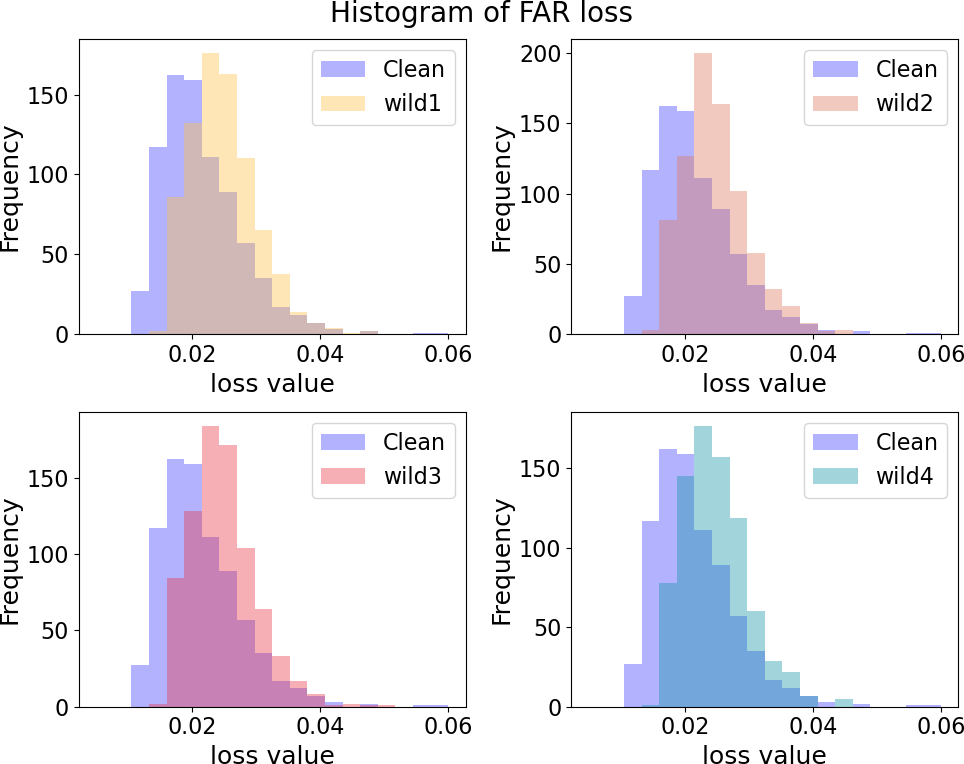}
    \caption{Histogram of FAR loss values on natural HR images (clean) and their versions with different degradations.}
    \vspace{-0.2cm}  
    \label{fig:abla_far}
\end{figure}

\noindent\textbf{Distribution shift revealed by FAR.}
FAR assumes the self-supervised task in \cref{eq:far} trained for natural images reveals the distribution shift when encountering images of lower quality. 
We validate this assumption by comparing the statistics of FAR values when inputs are natural images and their degraded versions. 
Specifically, we use the ground truth in the training set of DIV2K and its versions with different wild degradations~\cite{8575282}.
As shown in \cref{fig:abla_far}, FAR values generally increase when images have degradations, successfully revealing the distribution shift.

\noindent\textbf{Choice of pretrained image encoder.} In FAR, we choose the image encoder of CLIP~\cite{radford2021learning} as the encoder to be aligned with. \cref{tab:abla_Eclip} shows results of using the SR model EDSR~\cite{lim2017enhanced} and the classification model VGG19~\cite{simonyan2014very}. EDSR yields a high NRQM score but poor pixel-wise fidelity. The VGG model works but performs lower than $E_{clip}$. 
We select the CLIP encoder based on its good performance and easy use.
\begin{table}[h!]
    \centering
    \resizebox{0.85\linewidth}{!}{
    \begin{tabular}{c|c c c c}
    \hline
    \hline
      \multirow{2}{*}{Pretrained Encoder} & \multicolumn{4}{c}{Metrics} \\
    \cline{2-5}
         & PSNR & SSIM & LPIPS & NRQM \\
    \hline
    EDSR & 24.22&0.6878 &\Scd{0.2397} &\Bst{6.429} \\
    VGG &\Scd{25.55} &\Scd{0.7112} &0.2460 &6.102 \\
    $E_{clip}$ &\Bst{25.87} &\Bst{0.7204} &\Bst{0.2369} &\Scd{6.137} \\
    \hline
    \end{tabular}}
    \caption{Different choices of pretrained image encoder in FAR.}
    \label{tab:abla_Eclip}
\end{table}

\noindent\textbf{Strategies of generating controller $\mathbf{s}$} are different in pretraining and finetuning stages (see \cref{eq:s}).  
Here, we lower the extent of degradation for high HQI, \ie $\mathbf{s}=\mathbf{n}+(1-HQI(X^r, \mathrm{M}(X^r)))$, in the finetuning stage. 
As shown in \cref{tab:abla_s}, set degradation extent negatively proportional to HQI in the finetuning stage results in lower performances.
\begin{table}[h!]
    \centering
    \resizebox{\linewidth}{!}{
    \begin{tabular}{c|c c c c}
    \hline
    \hline
       $\mathbf{s}$  & PSNR & SSIM & LPIPS & NRQM \\
       \hline
        $\mathbf{n}+1-\text{HQI}(\mathrm{M}(X^r))$ &25.79 &0.7196 &0.237 &6.128 \\
        $\mathbf{n}+\text{HQI}(\mathrm{M}(X^r))$&\Bst{25.87} &\Bst{0.7204} &\Bst{0.2369} &\Bst{6.137} \\
    \hline
    \end{tabular}}
    \caption{Results of defining $\mathbf{s}$ differently in the finetuning. The strategy in \cref{eq:s} performs better.}
    \label{tab:abla_s}
\end{table}

\begin{figure}[h]
    \centering
    \InsertSubfig{0.49}{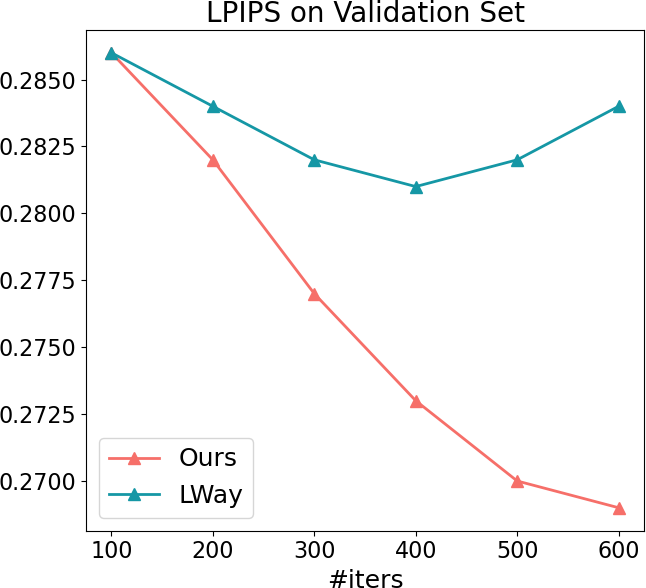}
    \InsertSubfig{0.48}{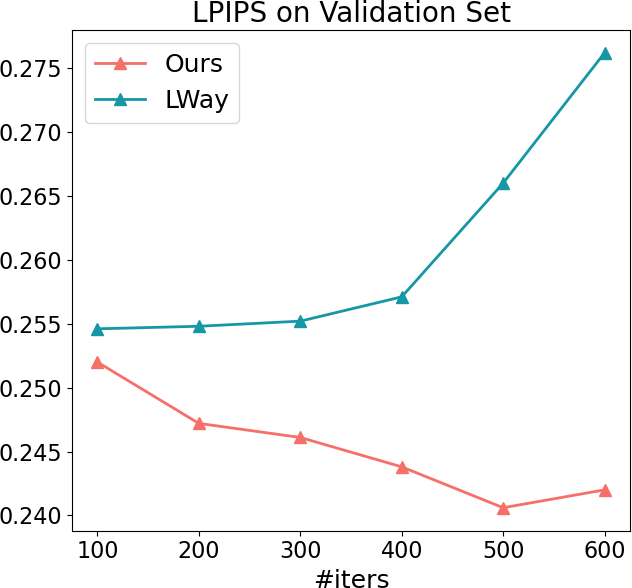}
    \vspace{-0.1cm}
    \caption{LPIPS scores changing with finetuning iterations. The validation sets are split from RealSR (left) and NTIRE20 (right).}
    \vspace{-0.2cm}
    \label{fig:iter}
\end{figure}

\noindent\textbf{Iterations of Finetuning.}
SR models can overfit the LR reconstruction objective.
In \cref{tab:quant_cmp}, for the reproduction of LWay~\cite{chen2024low} and our method, we split a validation set from the testing data and use the early stopping strategy. \cref{fig:iter} plots the LPIPS scores changing with finetuning iterations for LWay and our method. Models are trained with the same learning rate for 600 iterations. LWay can result in significant performance drops after 400 iterations for the two validation sets, while our method does not, likely due to the constraint from FAR. Our method is more appropriate than LWay when early stopping is infeasible. 

\section{Conclusion}
This paper studies self-supervised real-world image super-resolution by strengthening its awareness of high-resolution images. The core of our method involves a controller $\mathbf{s}$ and feature-alignment regularizer (FAR).
We consider two key limitations of existing self-supervised SR methods: reliance on insufficient degradation modeling and limited information from LR images, making the self-supervised objective not align well with super-resolution performance.

As such, the controller $\mathbf{s}$ enables adjustable degradation modeling based on super-resolution quality. FAR constrains the disparity between super-resolved and natural image distributions. 
Our method fintunes off-the-shelf SR models for specific real-world domains. Experiments show its effectiveness across multiple datasets, especially for important perceptual quality. 
Additionally, we provided directions for improving the self-supervised SR through finer degradation adjustments and improved prior constraints.
\clearpage
{
    \small
    \bibliographystyle{ieeenat_fullname}
    \bibliography{main}
}


\end{document}